\documentclass[12pt]{article}

\usepackage[dvips]{graphics,epsfig}
\usepackage[latin1]{inputenc}
\usepackage{setspace}
\usepackage{tabularx}
\usepackage{amsfonts}
\usepackage{booktabs}
\usepackage{bm}
\usepackage{cite}
\usepackage{textcomp}
\usepackage{amsmath}
\usepackage{graphicx}
\usepackage{float}
\usepackage{braket}
\usepackage{subfig}
\usepackage{dsfont}
\graphicspath{ {./plots/} }

\usepackage[unicode=true,bookmarks=false,breaklinks=false,pdfborder={0 0 1},colorlinks=true]
 {hyperref}
\hypersetup{
 citecolor=blue,linkcolor=blue,urlcolor=blue}

\def\laq{\raise 0.4 ex \hbox{$<$}\kern -0.8 em\lower 0.62 ex\hbox{$\sim$}}
\def\gaq{\raise 0.4 ex \hbox{$>$}\kern -0.7 em\lower 0.62 ex\hbox{$\sim$}}

\def\beq{\begin{equation}}
\def\eeq{\end{equation}}
\def\beqa{\begin{eqnarray}} 
\def\eeqa{\end{eqnarray}}
\begin{document}

\pagestyle{plain}

\begin{flushright}
{\bf DRAFT VERSION}\\
\today
\end{flushright}
\vspace{0.2cm}

\begin{center}

{\Large\bf A dynamic theory of entanglement for uniformly accelerated atoms}

\vspace*{0.5cm}

M. S. Soares\footnote{E-mail address: \url{matheus.soares@cbpf.br}}, N. F. Svaiter\footnote{E-mail address: \url{nfuxsvai@cbpf.br}}\\
{Centro Brasileiro de Pesquisas F\'{\i}sicas\\
Rua Xavier Sigaud, 150 - Urca, Rio de Janeiro - RJ, 22290-180, Brazil}\\

\vspace*{0.5cm}

G. Menezes\footnote{E-mail address: \url{gabriel.menezes10@unesp.br}}\\
{Instituto de F\'isica Te\'orica, Universidade Estadual Paulista\\ Rua Dr. Bento Teobaldo Ferraz, 271 - Bloco II, 01140-070 S\~ao Paulo, SP, Brazil}\\
\end{center}

\begin{abstract}
We study the entanglement dynamics of accelerated atoms using the theory of open quantum systems. We consider two atoms travelling along different hyperbolic trajectories with different proper times. We use the generalized master equation to discuss the dynamics of a pair of dipoles interacting with the electromagnetic field. The fundamental role played by proper acceleration in the entanglement harvesting and sudden death phenomenon is discussed. Finally, we discuss how the polarization of the atoms affects our results. For some choices of the polarization's configuration, we no longer observe the entanglement harvesting phenomenon.
\end{abstract}

\section{Introduction}

Entanglement remains a captivating property of quantum mechanics that has no analogue in classical physics. One definition is that it occurs when the Hilbert state of a closed system cannot be described as a tensor product of pure states of subsystems \cite{cohen1986quantum}. The long range correlation that appears has opened up new frontiers for many research areas such as quantum information, quantum computation and quantum optics \cite{nielsen2011}. In this work, we discuss the entanglement dynamics of uniformly accelerated atoms using a formalism constructed to deal with open quantum systems \cite{breuer2002theory}. With insights from the Unruh-Davies effect, we intend to merge some protocols of quantum optics and information \cite{Unruh:1976db, Davies:1974th,sciamacandelasdeutsch1981}.

In Minkowski spacetime, inertial observers may perform the canonical quantization with a Poincar\'e invariant Minkowski vacuum $\ket{0,M}$. How to implement the canonical quantization for observers in Minkowski spacetime that do not travel in inertial worldlines \cite{Bire82}? In 1973, Fulling discussed the quantization of a massive scalar field performed by an observer in a uniformly accelerated reference frame \cite{fulling1973}. Using the fact that the worldlines of these observers are integral curves of a timelike Killing vectors field, the canonical quantization can also be implemented in this situation. The ground state defined in this physical situation is the Fulling-Rindler vacuum $\ket{0,R}$. The fact that we are studying systems with infinitely many degrees of freedom lead us to non-unitarily equivalent quantization procedures. Instead of discussing this non-unitarily equivalent quantization using the Bogoliubov coefficients, a different route to shed some light on the problem of defining elementary excitations without the Poincaré group is to introduce a device constructed to detect such excitations. In the late 1970s, Unruh and Davies proved that an accelerated observer perceives the Minkowski vacuum as a thermal bath  \cite{Unruh:1976db, Davies:1974th}. They considered a model detector at rest in a uniformly accelerated reference frame interacting with a scalar field prepared in the Minkowski vacuum. They obtained that for the uniformly accelerated observer the Minkowski vacuum appears as a thermal bath of Rindler quasi-particles. It is worth noting that the Unruh-Davies effect also appears for uniformly accelerated observers using a Glauber detector \cite{soares21}. This topic has been widely discussed  \cite{Candelas:1976jv,svaiter1992,Ford:1994zz,Shih-Yuin:06,Crispino:2007eb,Lenz:2010vn,Arias:2011yg,Menezes:2015iva,Ben-Benjamin:2019opz,Rodriguez-Camargo:2016fbq, Arias:2015moa,Fewster:2016ewy,Menezes:2016quu, Menezes:2017rby, Picanco:2020api, PhysRevD.101.025009,Tjoa:2022oxv,kaplanek2020hot,arrechea2021inversion,carballo2019unruh,juarez2014onset}

Relativistic Quantum Information (RQI) has laid the theoretical foundations for understanding quantum phenomena in the context of accelerated reference frames and also for systems in a gravitational field governed by General Relativity \cite{Peres:2002ip,Gingrich:2002ota,Gingrich:2002otaa,Shi:2004yt,Czachor,Jordan:2006kt,Adesso:2007wi,Datta,Peres:2002wx,Lin:2010zzb,Bruschi:2010mc,Ostapchuk:2011ud,Hu:2012jr,CQG12,Lin:2015aua, Liu:2023lok}.  This framework helps us to understand entanglement dynamics generated by the Unruh-Davies effect. Here, we are interested in discussing entanglement harvesting and entanglement degradation for atoms travelling in different accelerated worldlines. The phenomenon that occurs when atoms initially in a non-separable state evolve to a separable state is known as \textit{entanglement degradation} \cite{alsing03, fuentes05}. On the other hand, in \textit{entanglement harvesting}, a pair of separable (or very weakly entangled) atoms is prepared to extract pre-existing entanglement from a quantum field \cite{Tjoa:2020riy, martinmartinez_prd21, Bhattacharya:2022ahn, Hu:2022nxc}. There are two ways to study entanglement harvesting and degradation.
One method uses a compactly supported switching function to compute the final density matrix of the atoms. The other method computes the density matrix (from the master equation) at a specific time $t$ by assuming a constant switching function. The latter is the one we employ in this work.

Kaplanek and Tjoa explore the use of open quantum systems in the sense of relativistic quantum information and found a relation of the timescale to assure that the results are reliable for non-Markovian processes \cite{tjoa23}. In fact, the procedure to derive the master equation used in Ref. \cite{soares23} and the one we use in this work is known as the \textit{microscopic derivation} and is guided through physical approximations \cite{breuer2002theory}. One of this approximations leads us to discuss what is \textit{late time}. For the atoms with the same proper acceleration, this discussion leads us to the same relation obtained in Ref. \cite{tjoa23} which assures that our master equation does not violate the non-Markovian processes limits.

The entanglement harvesting protocol is generally used to ensure that the entanglement is indeed harvested from the field and not due to any causal communication between the detectors, the protocol operates within short timescales. These timescales are short enough to prevent light-speed signals from being exchanged between the detectors, which would otherwise allow for signalling. It consists of considering atoms locally interacting with the same quantum field with an entangled ground state. This approach assures that entanglement of formation is provided by the field and we can study better the role of the quantum field to entanglement harvesting. Many interesting results were obtained with such a technique, see for example the Refs. \cite{She:2019hjv,He:2020xhz,Bhattacharya:2021zgd, martinmartinez-PhysRevD.92.064042,Sachs:2017exo,Pozas-Kerstjens:2017xjr,MM-PhysRevD.94.064074,Liu:2020jaj,Perche:2021clp, wu2023birth, wu2023accelerating, mann_2023correlation, martin24feb}. 

The study of entanglement dynamics in open quantum systems has illuminated how quantum correlations evolve when systems interact with their surrounding environments. These interactions introduce the notion of decoherence and can significantly impact the persistence of entanglement. Investigating the entanglement dynamics of uniformly accelerated atoms within this framework becomes particularly intriguing, as it allows us to explore how relativistic effects and open system dynamics combine to influence quantum correlations in complex ways \cite{benatti04_pra, hu2015, yu2016pra,hu2023}. In recent work we have derived a master equation to deal with qubits, interacting with a scalar field, in different reference frames with different proper times to explore the role of the state of motion in entanglement dynamics \cite{soares23}. In this work, we have demonstrated that different proper accelerations affect the entanglement dynamics more than other system variables such as the orthogonal distance of the qubits. This is the path we are going to follow in this work.

Here we generalize the results of Ref. \cite{soares23} for the case of atoms interacting with an electromagnetic field instead of the scalar field and also with atoms with different proper accelerations, generalizing the results of Ref. \cite{yu2016pra}. We revisit the entanglement dynamics of uniformly accelerated atoms using a generalized master equation. We use this procedure to investigate how acceleration influences the entanglement dynamics. By considering an initial non-separable state for the atoms and changing the acceleration and polarization conditions, we aim to shed light on the behavior and degradation of initial entangled states. We also want to explore the phenomena of entanglement harvesting by considering an initial separable state for the atoms and changing these two variables of our problem. In this case, we observe that for some polarization of the atoms we do not observe any creation of entangled states in some initial states of the system. We present an alternative view of entanglement dynamics for non-inertial atoms in a quantum optics system. 

The organization of this papers is as follows.  In Sec. \ref{sec:coupling_atoms} we present the formalism of the master equation for a pair of accelerated dipoles interacting with the electromagnetic field. We discuss the solutions of the derived master equation and study how the entanglement dynamics (entanglement degradation and the entanglement harvesting) is affected by the different values of both proper accelerations of the atoms in Sec. \ref{sec:ent_dyn}. Conclusions are given in sec. \ref{sec:conclusions}. We use unit such that $\hbar = c = k_{B} = 1$. Graphs were drawn using the Mathematica packages.
 \section{Master Equation for Accelerated Atoms}\label{sec:coupling_atoms}
\begin{figure}[htp!]
	\centering
	\includegraphics[scale=0.3]{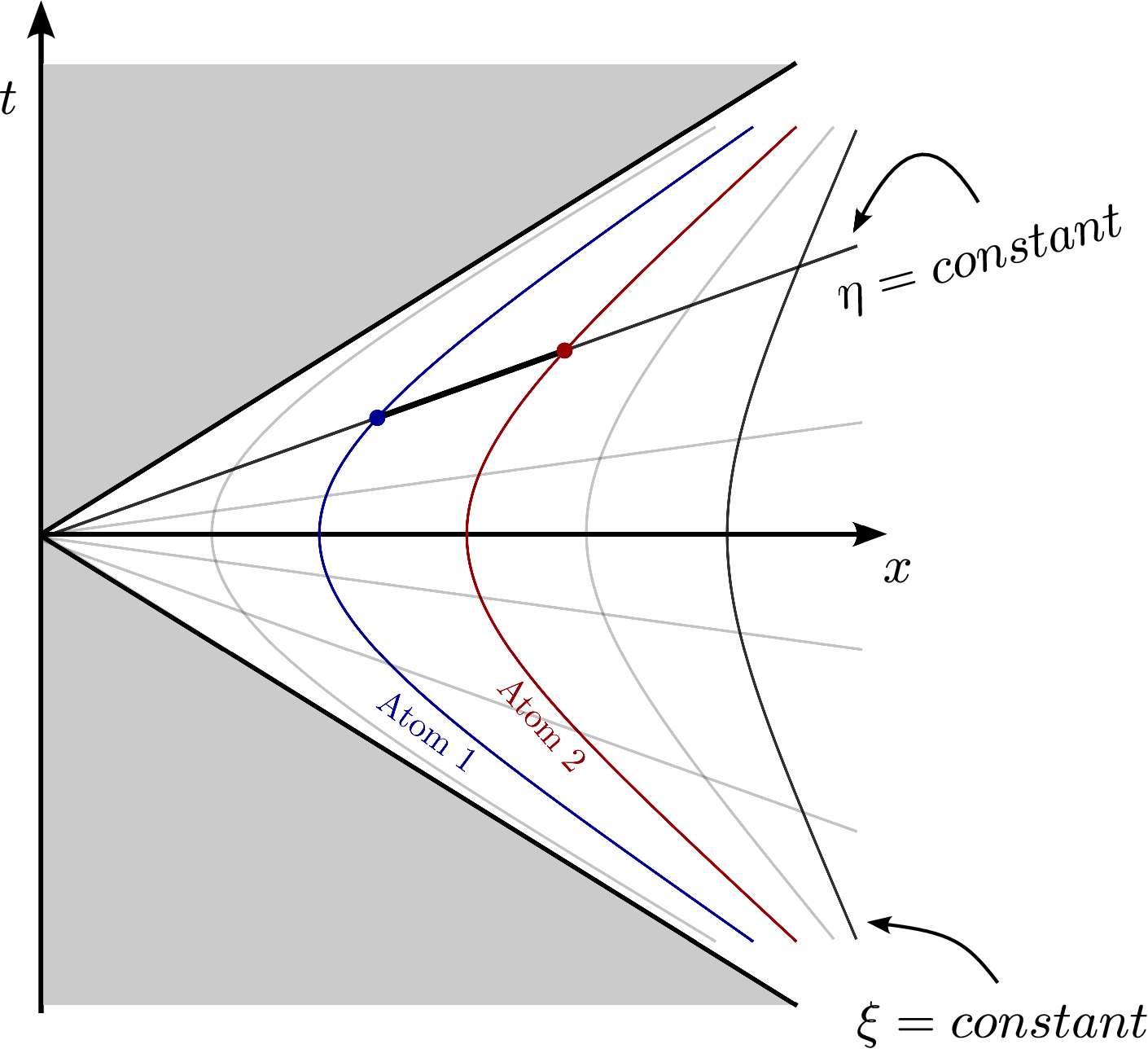}
	\caption{Spacetime diagram illustrating the Rindler Coordinates. The straight lines represent simultaneous hypersurfaces with constant $\eta$, while different values of $\xi$ represent different hyperbolae. The blue hyperbola represents the worldline of Atom 1, while the red one represents the worldline of Atom 2.}
	\label{fig:rindlercoord}
	\end{figure}

We start discussing radiative processes of atoms with two discrete energy levels: the ground state $\ket{g}$ and the excited state $\ket{e}$. Let us suppose that the two identical two-level atoms interact with a common quantized electromagnetic field. The atoms are moving along different hyperbolic trajectories in the $(t, x)$ plane and its coordinates of such motion is known as the Rindler coordinates $(\eta, \xi)$ and has the form
	\begin{eqnarray}
	x = \frac{e^{a \xi}}{a}\cosh a \eta,\nonumber\\
	t = \frac{e^{a \xi}}{a}\sinh a \eta,\label{eqq:rind_coord}
	\end{eqnarray}
where $a$ is a positive constant. The Rindler metric is $ds^2 = e^{2 a \xi} (d\eta^2 - d\xi^2) - d \mathbf{y}^2$, with $\mathbf{y}$ being the coordinates perpendicular to the motion of the atoms. The coordinates $(\eta, \xi)$ cover only a quadrant of Minkowski space, namely, the wedge $x > |t|$. Lines of constant $\eta$ are straight, whereas lines of constant $\xi$ are hyperbolas
	\begin{equation}
	x^{2} - t^2 = a^{-2}e^{2 a \xi},
	\end{equation}
representing world lines of uniformly accelerated observers with associated proper acceleration 
$ a e^{-a \xi}$. Hence different values for $\xi$ correspond to different hyperbolae and hence to different proper accelerations. The accelerated observer's proper time $\tau$ is related to $\xi, \eta$ by
	\begin{equation}
	\tau = e^{a \xi} \eta.\label{eqq:timeparameter}
	\end{equation}
We have to deal with two proper times since the atoms are moving along distinct hyperbolic trajectories. To derive the Heisenberg equations of motion of the coupled system, we have to choose a common time variable. To this purpose, we name the two atoms as Atom $1$ and Atom $2$. In Ref. \cite{soares23} we have used the proper time of the Atom $1$, $\tau_1$. In this work we use the Rindler coordinate time $\eta$ which gives us some technical simplifications to write the positive frequency Wightman two-point correlation function for the electrogmanetic field. Therefore, we describe the time evolution with respect to such a parameter, which, because of (\ref{eqq:timeparameter}), has a functional relation to each of the proper times of the atoms. We illustrate the trajectories of the atoms and the Rindler coordinates in Fig. \ref{fig:rindlercoord}. Once the time evolution is chosen to be $\eta$, by the relations given by Eq. (\ref{eqq:timeparameter}) both atoms are separated by a spacelike distance for each value of $\eta$.

The Hamiltonian that governs the time evolution of this atomic system with respect to $\eta$ can be written as
	\begin{equation}
	H_{A}(\eta) = \frac{\omega}{2}\left[\left[S^{Z}_{1}(\tau_{1}(\eta))\otimes \mathds{1}\right]\frac{d \tau_{1}}{d\eta}+\mathds{1}\otimes\left[S^{Z}_{2}(\tau_{2}(\eta))\right]\frac{d \tau_{2}}{d\eta}\right],\label{eqq:atomichamiltonian}
	\end{equation}
where $\omega$ is the energy gap of the atoms, $S^{Z}_{a} = \ket{e_{a}}\bra{e_{a}} - \ket{g_{a}}\bra{g_{a}}$ and the label $A$ is just a reference to the atomic system.

The mode decomposition of the field in terms of creation and annihilation operators can be obtained in the Coulomb gauge following the standard procedure of canonical quantization. In the Coulomb gauge, radiation fields are given by the vector potential $\textbf{A}$
	\begin{equation}
	\textbf{E} = - \frac{\partial \textbf{A}}{\partial t} \quad \text{and} \quad \textbf{B} = \nabla \times \textbf{A},
	\end{equation}
which is purely transverse. Because of such transversality, the interaction Hamiltonian in the proper time of the atoms can be written in a more convenient form which we will present in the following \cite{cohen1997photons}. Following, the fields can be written as
	\begin{eqnarray}
	  \textbf{E}(t, \textbf{x}) = \sum_{\lambda} \int d^3 \textbf{k} \frac{i \omega_{\textbf{k}}}{(2\pi)^3 (2 \omega_{\textbf{k}})^{1/2}}\left[a_{\textbf{k}, \lambda}(t) e^{i \textbf{k} \cdot \textbf{x}}- a^{\dagger}_{\textbf{k}, \lambda}(t) e^{-i \textbf{k} \cdot \textbf{x}}\right]\hat{\varepsilon}_{\lambda}(\textbf{k}),\label{eqq:Efielddecomp}\\
		  \textbf{B}(t, \textbf{x}) = \sum_{\lambda} \int d^3 \textbf{k} \frac{i \omega_{\textbf{k}}}{(2\pi)^3 (2 \omega_{\textbf{k}})^{1/2}}\left[a_{\textbf{k}, \lambda}(t) e^{i \textbf{k} \cdot \textbf{x}} - a^{\dagger}_{\textbf{k}, \lambda}(t) e^{-i \textbf{k} \cdot \textbf{x}}\right]\left[\hat{\textbf{k}}\times\hat{\varepsilon}_{\lambda}(\textbf{k})\right],\label{eqq:Bfielddecomp}
	\end{eqnarray}		
where $\omega_{\bm{k}} = k_{0}$ is the field mode frequency and $\hat{\varepsilon}_{\lambda}(\textbf{k})$ is the polarization's unit vector for each mode of the field that corresponds to the two polarization states of the photon. Also, the operators $a_{\textbf{k}, \lambda}(t)$ and $a^{\dagger}_{\textbf{k}, \lambda}(t)$  are defined by
\begin{eqnarray}
	a_{\textbf{k}, \lambda}(t) &=&  a_{\textbf{k}, \lambda}e^{ - i \omega_{\bm{k}} t}\\
	a^{\dagger}_{\textbf{k}, \lambda}(t) &=&  a^{\dagger}_{\textbf{k}, \lambda}e^{  i \omega_{\bm{k}} t}
\end{eqnarray}
where $a_{\textbf{k}, \lambda}$ and $a^{\dagger}_{\textbf{k}, \lambda}$ are the annihilation and creation operators of photons with polarization $\lambda$. 

The electromagnetic field Hamiltonian is written in the traditional form
	\begin{equation}
	H_{F} =  \sum_{\lambda}\int \frac{d^3 \textbf{k}}{(2\pi)^3} \omega_{\textbf{k}} a^{\dagger}_{\textbf{k}, \lambda}(t) a_{\textbf{k}, \lambda}(t).\label{eqq:emhamiltonian}
	\end{equation}
In the so-called electric dipole approximation of the multipolar coupling treatment one has that the Hamiltonian that describes the interaction between the atoms and the field is given by \cite{passante98, zhu2006, meystre2021quantum, agarwal2006quantum}
	\begin{eqnarray}
	  H_{I}(\eta) = - \boldsymbol{\mu}^{(1)}(\tau_{1}(\eta))\cdot \textbf{E}\left[x_{1}(\tau_{1}(\eta))\right]\frac{d\tau_{1}}{d\eta} -  \boldsymbol{\mu}^{(2)}(\tau_{2}(\eta))\cdot \textbf{E}\left[x_{2}(\tau_{2}(\eta))\right]\frac{d\tau_{2}}{d\eta},\label{eqq:HIeta}
	\end{eqnarray}

with the electric dipole moment operator for the $\alpha$-th atom $\boldsymbol{\mu}^{(\alpha)}(\tau_{\alpha}(\eta))$ given by
	\begin{equation}
		\boldsymbol{\mu}^{(\alpha)} = \mathbf{d}^{(\alpha)} S^{(\alpha)}_{-}e^{- i \omega \tau_{\alpha}(\eta)} + \mathbf{d}^{ (\alpha)*} S^{(\alpha)}_{+}e^{ i \omega \tau_{\alpha}(\eta)},\label{eqq:dipoleoperator}
	\end{equation}
where we have defined the $\alpha$-th dipole raising and lowering, respectively, operators as $S^{(\alpha)}_{+} = \ket{e_{\alpha}}\bra{g_{\alpha}}$ and $S_{-}^{(\alpha)} = \ket{g_{\alpha}}\bra{e_{\alpha}}$ and $\mathbf{d}^{(\alpha)} = \bra{g_{\alpha}} \boldsymbol{\mu}^{(\alpha)} \ket{e_{\alpha}} $ is the transition matrix element of the dipole moment operator for the $\alpha$-th atom. To avoid any possible notation misleading we use Greek indices $(\alpha, \beta = 1$ or $2$) to describe the atoms and Latin indices ($i, j = 1, 2$ or $3$) to describe vector's components. Also, we are not considering the summation convention over repeated indices in this work. 

The factor $d\tau_{\alpha}/d\eta = e^{a \xi_{\alpha}}\equiv a_{\alpha}$ which appears in Eq. (\ref{eqq:HIeta}) is justified because we are considering that both atoms are travelling along different stationary trajectories and will play a crucial role in the generalized master equation.


The procedure to obtain a master equation for atoms at rest in non-inertial frames can be realized via the \textit{microscopic derivation} applying some techniques of quantum field theory in curved space-time. This was obtained in Refs. \cite{benatti04_pra, hu2015} but we recall that in this work we are considering that the atoms are travelling along different hyperbolas and a generalization of the master equation obtained in the mentioned works is required. See for example Ref. \cite{soares23}. To obtain such master equation for the system we are interested in this work we are making some identifications. The master equation obtained in the mentioned work has the standard Lindblad form
	\begin{equation}\label{eqq:gme_final_form}
	\dot{\rho}_{A}(\eta) = - i \left[H_{eff},\rho_{A}(\eta)\right] + \mathcal{L}\left\lbrace\rho_{A}(\eta)\right\rbrace,
	\end{equation}	
with the effective Hamiltonian defined by
	\begin{equation}\label{eqq:h_eff}
	H_{eff} \equiv H_{A} + \sum_{\nu = -\omega}^{\omega}\sum_{i,j}\sum_{\alpha, \beta}\Delta^{(\alpha \beta)}_{ij}(a_{\beta} \nu)A^{\dagger (\alpha)}_{i}(\nu) A^{(\beta)}_{j}(\nu).
	\end{equation}
The term, $\mathcal{L}\left\lbrace\rho_{A}(\eta)\right\rbrace$, usually denoted as the dissipative contribution of the generalized master equation, is given by
	\begin{eqnarray}
	\mathcal{L}\left\lbrace\rho_{A}(\eta)\right\rbrace = \frac{1}{2}\sum_{\nu = -\omega}^{\omega}\sum_{i,j}\sum_{\alpha, \beta} \Gamma^{(\alpha \beta)}_{ij}(a_{\beta} \nu)\left( 2 A^{(\beta)}_{j}(\nu)\rho_{A}(\eta) A^{\dagger (\alpha)}_{i}(\nu)\right.\nonumber\\
	-\left.\left\lbrace A^{\dagger (\alpha)}_{i}(\nu) A^{(\beta)}_{j}(\nu),\rho_{A}(\eta)\right\rbrace\right),\label{eqq:dissipative_term}
	\end{eqnarray}
where we have associated the atomic operators as
	\begin{eqnarray}
	A^{(\alpha)}_{i}(\omega) &=& d_{i}^{(\alpha)}S_{-}^{(\alpha)},\\
 A^{(\alpha)}_{i}(-\omega) &=& A^{\dagger (\alpha)}_{i}(\omega) =  d_{i}^{*(\alpha)}S_{+}^{(\alpha)}.\label{eqq:lindbaldops}
	\end{eqnarray}	
The $\Gamma^{(\alpha \beta)}_{ij}$, $\Delta^{(\alpha \beta)}_{ij}$ are, respectively, real and complex functions
	\begin{eqnarray}
	\Gamma^{(\alpha \beta)}_{ij}(a_{\beta} \nu) =  W_{ij}^{(\alpha \beta)}(a_{\beta} \nu) +  W_{ji}^{*(\beta \alpha)}(a_{\beta} \nu),\label{eqq:gamma_gen}\\
	\Delta^{(\alpha \beta)}_{ij}(a_{\beta} \nu) = \frac{W_{ij}^{(\alpha \beta)}(a_{\beta} \nu) -  W_{ji}^{*(\beta \alpha)}(a_{\beta} \nu)}{2i}, \label{eqq:Delta_gen}
\end{eqnarray}	
where the function $W_{ij}^{(\alpha \beta)}(a_{\beta} \nu)$ is obtained by an integral transform of the electromagnetic field correlation functions which will be defined in the following. 

For various systems we are interested in, these correlation functions can usually be written as functions of the difference of the time parameter $\eta - \eta'$. We then use  $x_{\alpha}(\tau_{\alpha}(\eta)) = x_{\alpha}(\eta)$ to simplify our notations and write the electromagnetic field correlation functions as
    \begin{equation}\label{eqq:wightman_def}
        G^{(\alpha \beta)}_{ij}(\eta - \eta') = \bra{0,M} E_{i}( x_{\alpha}(\eta))E_{j}( x_{\beta}(\eta'))\ket{0,M},
    \end{equation}
where $\ket{0,M}$ is the Minkowski vacuum.
    
Using that $s = \eta - \eta'$ and Eq. 
 (\ref{eqq:timeparameter}), the $W_{ij}^{(\alpha \beta)}(a_{\beta} \omega)$ function can be written as the following
    \begin{eqnarray}
         W_{ij}^{(\alpha \beta)}(a_{\beta} \nu) = a_{\alpha} a_{\beta}\int_{0}^{\infty} ds~ e^{i a_{\beta} \nu s }G^{(\alpha \beta)}_{ij}(s).\label{eqq:wij}
    \end{eqnarray}
It is important to point out that depending on the functional form of $G^{\alpha \beta}_{ij}(s)$ the above equation is not well defined since the singularity of the integrand coincides at the lower limit of the integral. What is well defined is the the combination given by Eq. (\ref{eqq:gamma_gen}). 

The electromagnetic field correlation functions for inertial observers with the field prepared in the Minkowski vacuum is known as \cite{takagi1986vacuum}
    \begin{equation}
         \bra{0,M} E_{i}( t, \bm{x})E_{j}( t', \bm{x}')\ket{0,M} = -\frac{1}{4 \pi^2}(\partial_{0} \partial_{0}'\delta_{ij} - \partial_{i}\partial_{j}')\left(\frac{1}{(t - t' - i\epsilon)^2  - \Delta \mathbf{x}^2 }\right),\label{eqq:cor_func_inertial}
    \end{equation}
where $\Delta \mathbf{x}^2 = \Delta x^2 + \Delta y^2 + \Delta z^2$ and $\Delta x_{k}= x_{k} - x_{k}'$. To study the role of the Unruh-Davies effect in the entanglement dynamics, we need to write the correlation functions defined by Eq. (\ref{eqq:cor_func_inertial}) in terms of the Rindler coordinates from Eq. (\ref{eqq:rind_coord}). With a suitable coordinate transformation, we obtain:
    \begin{eqnarray}
    G^{(\alpha \beta)}_{ij}(s) = \frac{a^{4} \delta_{ij}}{4 \pi^{2} a_{\alpha}^{2} a_{\beta}^{2} }\frac{\nu_{i}^{(\alpha \beta)}(s)}{\left[\cosh \left(as - i  \epsilon\right) - \cosh\phi\right]^{3}},\label{eqq:corre_final_form}
    \end{eqnarray}
with 
    \begin{eqnarray}
        \nu_{1}^{(\alpha \beta)}(s) &=& \cosh\left(as - i  \epsilon\right) - \cosh\phi,\\
        \nu_{2}^{(\alpha \beta)}(s) &=& \nu_{3}^{(\alpha \beta)}(s) =\cosh\left(as - i  \epsilon\right) \cosh\phi -1,
    \end{eqnarray}
and
    \begin{equation}
        \cosh\phi = 1 + \frac{(a_{\alpha} - a_{\beta})^{2}}{2a_{\alpha}a_{\beta}}.
    \end{equation}
Using Eq. (\ref{eqq:corre_final_form}) in Eq. (\ref{eqq:wij}) and after some manipulations, we obtain the function $\Gamma_{ij}^{(\alpha \beta)}$ as the following Fourier transformation
	\begin{eqnarray}
         \Gamma_{ij}^{(\alpha \beta)}(a_{\beta} \nu) = a_{\alpha} a_{\beta}\int_{-\infty}^{\infty} ds~ e^{i a_{\beta} \nu s }G^{(\alpha \beta)}_{ij}(s).\label{eqq:gammaij}
    \end{eqnarray}

Therefore, we can write the dissipative contribution to the master equation in a more useful form using the operators $A^{\alpha}_{i}(\omega)$ of Eqs. (\ref{eqq:lindbaldops})
	\begin{eqnarray}
		\mathcal{L}\left\lbrace\rho_{A}(\eta)\right\rbrace =  \frac{1}{2} \sum_{\alpha, \beta = 1}^{2}\sum_{i,j = 1}^{3}C_{ij}^{(\alpha \beta)} \left[ 2S_{j}^{(\beta)}\rho(\eta)S^{(\alpha)}_{i} - \left\lbrace S^{(\alpha)}_{i}S_{j}^{(\beta)},\rho(\eta) \right\rbrace \right], \label{eqq:gme_em_field}
	\end{eqnarray}
where $S_{j}^{(1)} = \sigma_{j}\otimes\sigma_{0}$ and $S_{j}^{(2)} = \sigma_{0}\otimes\sigma_{j}$, where $\sigma_{j}$ being the $j$-th Pauli matrix. The Kossakowski matrix $C_{ij}^{(\alpha \beta)}$ is defined as
	\begin{equation}\label{kossa-matrix}
	C_{ij}^{(\alpha \beta)} = \delta_{ij} A^{(\alpha \beta)} - i\epsilon_{ij3} B^{(\alpha \beta)} - \delta_{i3}\delta_{3j}A^{(\alpha \beta)},
	\end{equation}
with 
	\begin{eqnarray}
		A^{(\alpha \beta)} &=& \frac{\Gamma^{(\alpha \beta)}(a_{\beta} \omega) + \Gamma^{(\alpha \beta)}(-a_{\beta} \omega)}{4},\\
		B^{(\alpha \beta)} &=& \frac{\Gamma^{(\alpha \beta)}(a_{\beta} \omega) - \Gamma^{(\alpha \beta)}(-a_{\beta} \omega)}{4},			
	\end{eqnarray}
and

	\begin{equation}
	\Gamma^{(\alpha \beta)}(a_{\beta} \omega) = \sum_{i, j} \Gamma^{(\alpha \beta)}_{ij}(a_{\beta} \omega)d^{* (\alpha)}_{i}d^{(\beta)}_{j}.\label{eqq:gamma_compact}
	\end{equation}

In \ref{app:gammas} we have obtained the explicit form of the functions $\Gamma_{ij}^{(\alpha \beta)}$. We then write those results defining $\varsigma_{\beta}$, a parameter related with the inverse of the proper acceleration of the atoms and with the energy gap $\omega$, as
	\begin{equation}\label{eqq:inv_acel}
	\varsigma_{\beta} = \frac{a_{\beta} \omega}{a}.
	\end{equation}
Then, from Eq. (\ref{eqq:gamma_compact}), the function $\Gamma^{(\alpha \beta)}(\varsigma_{\beta})$ for $\alpha = \beta$ is written as
	\begin{eqnarray}
		\Gamma^{(\beta)}(\varsigma_{\beta}) &= \frac{4}{3} \Gamma_{0} \left(\frac{1 + \varsigma_{\beta}^2}{\varsigma_{\beta}}\right)  \frac{1}{1 - e^{-2\pi \varsigma_{\beta}}}, \label{eqq:gamma_aa} 
	\end{eqnarray}
and for $\alpha\neq\beta$ 
		\begin{eqnarray}
		\Gamma^{(\alpha \beta)}(\varsigma_{\beta}) =  \Gamma_{0}\frac{{csch}^2\phi}{ \varsigma_{\alpha}\varsigma_{\beta}}    \sum_{i = 1}^{3}  \frac{\hat{d}_{i}^{* (\alpha)}\hat{d}_{i}^{(\beta)}\mathcal{U}_{i}(\varsigma_{\beta})}{ 1 - e^{- 2 \pi \varsigma_{\beta}}},\label{eqq:gamma_ab}
	\end{eqnarray}
where we have used $d_{i}^{\beta} = |\textbf{d}^{(\beta)}| \hat{d}_{i}^{(\beta)}$, with $\hat{d}_{i}^{(\beta)}$ being the unit vector of the dipole operator for the $\beta$-th atom. In this work  we are assuming that the magnitudes of the eletric dipoles of the atoms are the same, $i.e.$,  $|\bm{d}^{(1)}| = |\bm{d}^{(2)}| = |\bm{d}|$, but the orientations might be different. In Eq. (\ref{eqq:gamma_aa}) and \ref{eqq:gamma_ab} we have defined the parameter
	\begin{equation}
	\Gamma_{0} = \frac{a|\textbf{d}|^2 \omega^{2}}{4 \pi}.
	\end{equation}
and the function $\mathcal{U}_{i}(\varsigma_{\beta})$ defined by
	\begin{eqnarray}
	\mathcal{U}_{1}(\varsigma_{\beta}) &= 4\left[\coth\phi \sin\left(\varsigma_{\beta} \phi\right)-\varsigma_{\beta}\cos\left(\varsigma_{\beta} \phi\right) \right],\\
	\mathcal{U}_{2}(\varsigma_{\beta}) &= \mathcal{U}_{3}(\varsigma_{\beta})= 2 \left[ \varsigma_{\beta} \cos\left(\varsigma_{\beta} \phi\right)\cosh \phi \right. \nonumber\\ &\left.+ \sin\left(\varsigma_{\beta} \phi\right)\sinh \phi \left(\varsigma_{\beta}^2 -  {csch}^2 \phi\right) \right].
	\end{eqnarray}

We can construct the $A(B)^{(\alpha \beta)}$ functions that appears in the Kossakowski matrix of Eq. (\ref{kossa-matrix}). The results of this functions is computed for $\alpha = \beta$ as
	 \begin{eqnarray}
	 A^{(\beta \beta)} &=& \Gamma_{0} \frac{1 + \varsigma_{\beta}^2}{3\varsigma_{\beta}}\coth\left(\pi \varsigma_{\beta}\right),\label{eqq:aii-function}\\
	 B^{(\beta \beta)} &=& \Gamma_{0} \frac{1 + \varsigma_{\beta}^2}{3\varsigma_{\beta}},\label{eqq:bii-function}
	 \end{eqnarray}
and for $\alpha \neq \beta$
	\begin{eqnarray}
	A^{(\alpha \beta)} &=&  \Gamma_{0}\frac{ {csch}^2 \phi}{4 \varsigma_{\alpha}\varsigma_{\beta}} \coth\left( \pi \varsigma_{\beta}\right) \sum_{i = 1}^{3} \hat{d}_{i}^{* (\alpha)}\hat{d}_{i}^{(\beta)} \mathcal{U}_{i}(\varsigma_{\beta}) ,  \label{eqq:aij-function}\\
	B^{(\alpha \beta)} &=& \Gamma_{0}\frac{ {csch}^2 \phi}{4 \varsigma_{\alpha}\varsigma_{\beta}} \sum_{i = 1}^{3} \hat{d}_{i}^{* (\alpha)}\hat{d}_{i}^{(\beta)} \mathcal{U}_{i}(\varsigma_{\beta}). \label{eqq:bij-function}
\end{eqnarray}

As can be seen by Eq. (\ref{eqq:gme_em_field}), the terms responsible for changing the entanglement dynamics with respect to atom's acceleration and polarization are the functions $A^{(\alpha\beta)}$ and $B^{(\alpha\beta)}$ presented in Eqs. (\ref{eqq:aii-function}-\ref{eqq:bij-function}). The choice of the atom's proper acceleration is the same of changing $\varsigma_{\beta}$ and it is clear that this change affects these functions. The choice of atom's polarization implies a choice of the component of the functions $\mathcal{U}_{i}(\varsigma_{\beta})$ for $\alpha \neq \beta$ and also affects significantly the dynamics of the system.  This is the reason that we are going to study the entanglement dynamics for different atom's polarization and proper acceleration with the numerical solution of our master equation.  This will be discussed in the next section.

\section{Entanglement Dynamics}\label{sec:ent_dyn}

Once the master equation is obtained, we must analyze the solutions of its matrix elements $\rho_{ij}$ to study the entanglement dynamics. To this purpose we can use a function of its elements to determine how much entanglement exists in the system. There are many proposed functions to this purpose as the entanglement entropy, concurrence, negativity between others. In this work we choose to use the concurrence, introduced by Wooters in Ref. \cite{wootters98}.  
 
 To simplify this discussion and the calculations, we treat the two atoms' system as a four-level single system by working in the basis of the collective states
    \begin{eqnarray}
        \ket{G} &=& \ket{g_1} \otimes \ket{g_2}, \nonumber \\
        \ket{E} &=& \ket{e_1} \otimes \ket{e_2},\nonumber \\
        \ket{S} &=& \frac{1}{2}\left(\ket{g_1} \otimes \ket{e_2} + \ket{e_1} \otimes \ket{g_2}\right)\nonumber \\
        \ket{A} &=& \frac{1}{2}\left(\ket{g_1} \otimes \ket{e_2} - \ket{e_1} \otimes \ket{g_2}\right),  \label{eqq:colle_states}      
    \end{eqnarray}
where we have two separated states $\ket{G}$ and $\ket{E}$ and two maximally entangled states, the symmetric state $\ket{S}$ and the antisymmetric state $\ket{A}$. The reduced density matrix $\rho(\tau)$ in this basis can be written as a sum of a block-diagonal form and a off-diagonal block, $\rho_{off}$
	\begin{equation}\label{eqq:rho_initial_basis}
	\rho(\tau) = \left(\begin{array}{cccc}
	\rho_{GG} & \rho_{GE} & 0 & 0\\
	\rho_{EG} & \rho_{EE} & 0 & 0\\
	0 & 0 & \rho_{SS} & \rho_{SA}\\
	0 & 0 & \rho_{AS} & \rho_{AA}
	\end{array}\right) + \rho_{off}(\tau),
	\end{equation}
with $\rho_{ij} = \bra{i}\rho\ket{j}$. In our previous work we argue that the time evolution of the master equation preserves the block-diagonal form, therefore, by considering an initial block-diagonal density matrix we can disregard the off-diagonal contributions of Eq. (\ref{eqq:rho_initial_basis}) \cite{soares23}. In Ref. \cite{tjoa23} the authors prove that the off-diagonal contribution vanishes in the late-times regimes anyway. By these discussions we only consider the evolution of block-diagonal of Eq. (\ref{eqq:rho_initial_basis}).

As we are interested to discuss entanglement dynamics, the unitary evolution contribution of the right-hand side of Eq. (\ref{eqq:gme_final_form}) can be disregarded. The matrix elements of the reduced density matrix is computed using Eq. (\ref{eqq:gme_em_field}) in the basis of Eq. (\ref{eqq:colle_states}). We obtain eight coupled linear differential equations involving all the matrix components of Eq. (\ref{eqq:rho_initial_basis}).

When dealing with atoms exhibiting distinct proper accelerations, the matrix elements  $\rho_{AS}$ and $\rho_{SA}$ exhibit a more involved expression compared to those associated with atoms having the same proper acceleration, as demonstrated in Ref. \cite{yu2016pra}. Consequently, these matrix elements play an essential role in the entanglement measures. To study the entanglement dynamics we use the \textit{concurrence} \cite{wootters98}. The concurrence $C[\rho]$ is a function of the matrix elements of $\rho(\eta)$ which has a value $C = 0$ for separated states and $C = 1$ for maximally entangled states. It can be shown that the concurrence in this basis has the form
    \begin{equation}\label{eqq:concurrence}
        C[\rho] = max \left\lbrace 0, Q(\eta)\right\rbrace,
    \end{equation}
where $Q(\eta)$ is written as
    \begin{eqnarray}
          Q(\eta) = \sqrt{\left[\rho_{AA}(\eta) - \rho_{SS}(\eta) \right]^2 - \left[\rho_{AS}(\eta) - \rho_{SA}(\eta) \right]^2 }- 2 \sqrt{\rho_{GG}(\eta)\rho_{EE}(\eta)}.
    \end{eqnarray}

Before we solve the master equation given by Eq. (\ref{eqq:gme_final_form}) we perform an approximation of the values of the parameters $\varsigma_{\beta}$. We consider that $\varsigma_{2} >> \varsigma_{1}$ and $\varsigma_{1} << 1$. Note that for a fixed value of $\omega$, the choice of $\varsigma_{\beta}$ is the choice of hyperbolae from Fig. \ref{fig:rindlercoord} and the configuration $\varsigma_{1} < \varsigma_{2}$ is represented by choosing the blue hyperbola always on the left of the red one.  In this sense the functions of Eqs. (\ref{eqq:aii-function}-\ref{eqq:bij-function}) have the simplification: $A^{(2)} = B^{(2)}$, $A^{(12)} = B^{(12)}$ and $A^{(21)} = B^{(21)}$. The dissipation contribution explicitly presented in \ref{app:gmeterms} of this work already takes these approximations into account.

\subsection{Entanglement Sudden Death}

We start setting the atoms to initial entangled states. In this case, we observe the degradation of such states due to the interaction with the field. When this degradation occurs in a finite time time interval, this phenomenon is called  \textit{entanglement sudden death} \cite{esd09}. In the basis of Eq. (\ref{eqq:colle_states}) we can set the system to be on the initial state $\ket{S}$ or $\ket{A}$, where the concurrence will have the value $1$. To investigate the impact of atoms' proper acceleration on the entanglement dynamics, we pursue two distinct approaches. We fix one of the proper acceleration, $i.e.$, we fix $\varsigma_{2}$, and vary the other ($\varsigma_{1}$). Another approach we adopt involves varying both proper accelerations, that is, altering both $\varsigma_{1}$ and $\varsigma_{2}$. This latter approach results from the generalization undertaken in this study and the one presented in Ref. \cite{soares23}.

	\begin{figure}[htpb!]
	\centering
	\subfloat[]{\includegraphics[height=2in]{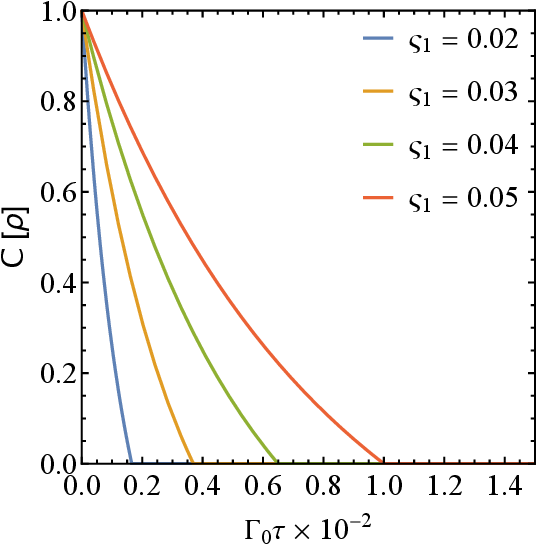}\label{deaths10}}\hspace*{2pt}
	\subfloat[]{\includegraphics[height=2in]{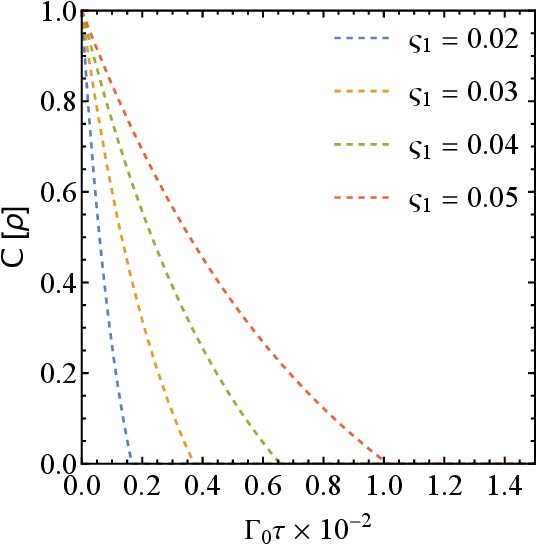}\label{deaths01}}\hfill
	\caption{ Comparison of entanglement sudden death for different values of $\varsigma_{1}$. The initial state were chosen to be $\ket{S}$ with $\varsigma_{2} = 0.6$. In (a), the polarization of the atoms is defined along the $x$ direction and for figure (b) is along the $z$ direction.}
	\end{figure}

In Fig. \ref{deaths10}, by considering the initial state as $\ket{S}$ and both atoms polarized along the $x$ direction, $\hat{d}^{(1)} =  \hat{d}^{(2)} = (1,0,0)$, we fix a value for $\varsigma_2$ and vary $\varsigma_{1}$. We observe that increasing $\varsigma_1$ the lifetime of the entangled states also increases. In fact, as $\varsigma_{\beta}$ is related to the \textit{inverse} of the proper acceleration, as defined in Eq. (\ref{eqq:inv_acel}), we have that decreasing the proper acceleration of one atom leads to the increase of the lifetime of entanglement. This relation of the degradation of entanglement and the proper acceleration of the atoms is due to the Unruh-Davies effect and is in agreement with previous results (with similar systems, but with atoms having the same proper acceleration) for the scalar field \cite{benatti04_pra, hu2015} and for the electromagnetic field \cite{yu2016pra}.

\begin{figure}[htpb!]
		\centering
		\includegraphics[height=2in]{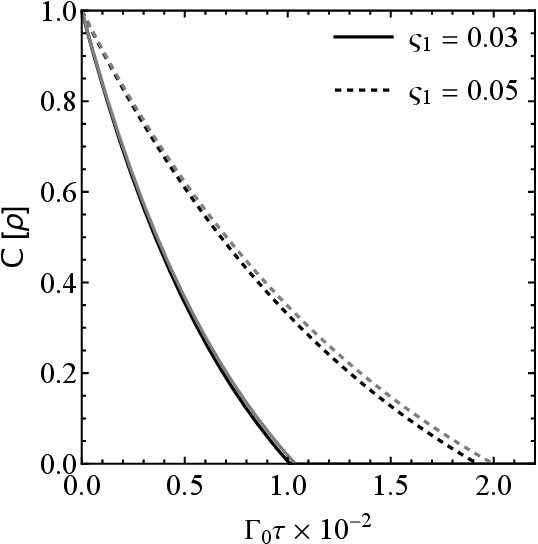}
		\caption{Comparison of entanglement degradation for atoms polarized along different directions. The black line corresponds to the polarization in the $x$ direction and the gray one corresponds to the $z$ direction. The initial state were chosen to be $\ket{S}$ with $\varsigma_{2} = 0.6$.}
		\label{death_comp}	
	\end{figure}

In Fig. \ref{deaths01} we consider the same system with the same initial conditions but with the atoms polarized along the $z$ axis, $\hat{d}^{(1)} =  \hat{d}^{(2)} = (0,0,1)$. Although the results for the two atom's polarization has the same pattern, in Fig. \ref{death_comp} we observe that the effects of polarization in this case is a subtly change in the lifetime of the entangled states, with the polarization along the $z$ axis being the one with the larger lifetime.

\begin{figure}[htpb!]
	\centering
	\subfloat[]{\includegraphics[height=2in]{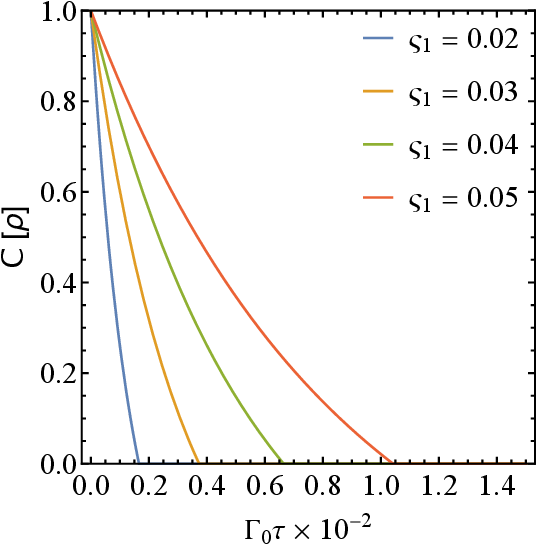}\label{deatha10}}\hspace*{2pt}
	\subfloat[]{\includegraphics[height=2in]{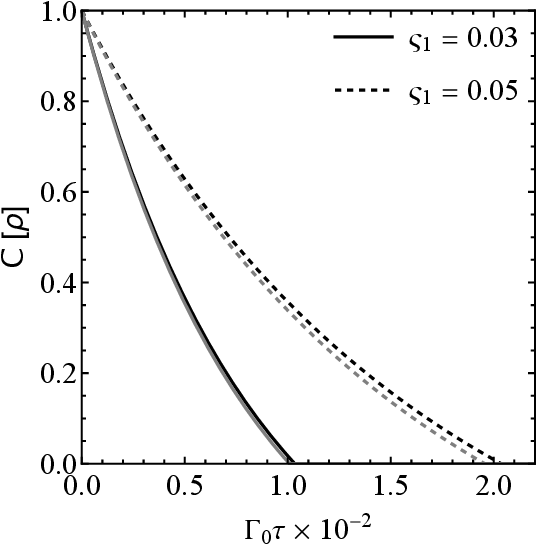}\label{deatha0110}}	
	\caption{ Comparison of entanglement sudden death for different values of $\varsigma_{1}$. The initial state were chosen to be $\ket{A}$ with $\varsigma_{2} = 0.6$. In (a), the polarization of the atoms is defined along the $x$ direction. In figure (b) we compare the entanglement sudden death for different values of polarization of the atoms. The black line corresponds to the polarization in the $x$ direction and the gray one corresponds to the $z$ direction.}
	\end{figure}

The dynamics appears in a similar way when considering the initial state of the system as the antissimetric one, $\ket{A}$. In Fig. \ref{deatha10} we observe that for a fixed value of $\varsigma_{2}$ and the atoms polarized along the $x$ direction, the lifetime of the entangled state decreases if the proper acceleration of the first atom increases. Compared with the situation presented in Fig. \ref{deaths10}, we observe an increase of the lifetime of the entangled states. For the atoms' polarized along the $z$ axis, we observe in Fig. \ref{deatha0110} that this choice, unlike the case for the initial state $\ket{S}$, is the one with larger entanglement lifetime.

\begin{figure}[htpb!]
\centering
\includegraphics[height=2in]{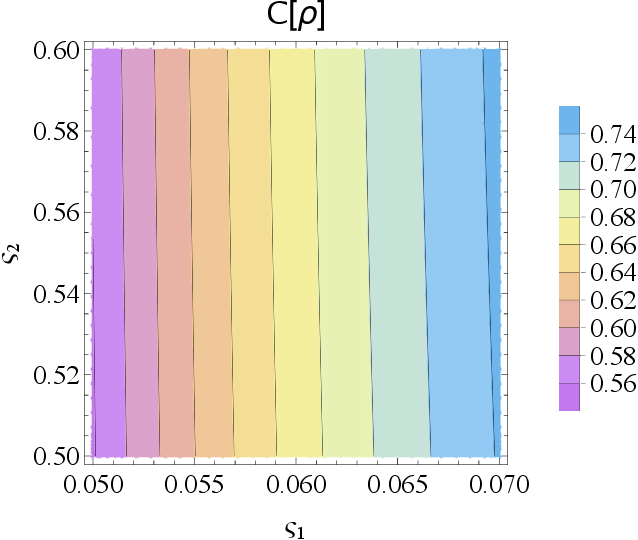}
\caption{ Comparison of entanglement sudden death for different values of $\varsigma_{1}$ and $\varsigma_{2}$. The initial state were chosen to be $\ket{S}$ with $\Gamma_{0}\tau = 0.003$ and the polarization of the atoms along the $x$ direction.}
\label{deathsboth}
\end{figure}

From the previous results, we observe that for the entanglement sudden death, the proper accelerations of the atoms play an important role for the dynamics. We observe that changes in the atoms' polarization only subtly affect the lifetime of the entangled states. As we have frequently mentioned in this work, the alternative possibility of the generalized master equation is to study the entanglement dynamics when the acceleration of \textit{both} atoms changes. From Ref. \cite{soares23}, we know that this fact produces an impact in the concurrence since the differential equations for the terms $\rho_{AS}$ and $\rho_{SA}$ do not have vanishing solutions. This results appears in a similar way for the system studied in this work as can be seen in Fig. \ref{deathsboth}, where we have fixed value of the time parameter $\Gamma_{0} \tau$, set both atoms with the polarization along the $x$ direction and vary the two parameters $\varsigma_1$ and $\varsigma_2$.  
        \begin{table*}[htpb!]
    \caption{Comparison of the entanglement degradation for uniformly accelerated atoms interacting with different type of Fields.}
    \label{comptable-death}
    \begin{tabular}{  l |  p{5cm} | p{5cm} }
        \toprule
     
& \textbf{Same proper acceleration's atoms}   
& \textbf{Different proper accelerations' atoms} \\\midrule
\textbf{Scalar Field} 
& the lifetime of the entangled states decreases as the proper acceleration increases \cite{benatti04_pra,hu2015, landulfo09}.   
& the lifetime of the entangled states decreases as the proper acceleration of one atom increases keeping the other's acceleration fixed \cite{soares23}. \\\hline
\textbf{Electromagnetic Field}     
& the lifetime of the entangled states decreases as the proper acceleration increases \cite{doukas09}. The choice of the atoms' polarization does not significantly change this dynamics \cite{yu2016pra}.                         
& for a fixed proper acceleration of one atom, the lifetime of the entangled states decreases as the proper acceleration of the other atom increases. The choice of the atoms' polarization does not significantly change this dynamics. \\\hline
\textbf{Dirac Field }       
& to be investigated. 
& to be investigated.  \\
        \bottomrule
    \end{tabular}
    \end{table*}

As seen in Fig. \ref{deathsboth}, fixing the parameter $\varsigma_{1}$ does not significantly affect the concurrence with variations in $\varsigma_{2}$. However, when $\varsigma_{2}$ is fixed, the concurrence changes significantly with variations in $\varsigma_{1}$. This result highlights the interesting scenario where both atoms can experience different proper accelerations and its impact for the entanglement dynamics. 

 In Table \ref{comptable-death}, we summarize the differences/similarities between our results for entanglement sudden death with the results for entanglement dynamics in non-inertial frames for different quantum fields found in the literature \cite{benatti04_pra,hu2015, landulfo09,soares23, doukas09, yu2016pra}.


\subsection{Entanglement Harvesting}

\begin{figure}[htpb!]
	\centering
	\subfloat[]{\includegraphics[height=2in]{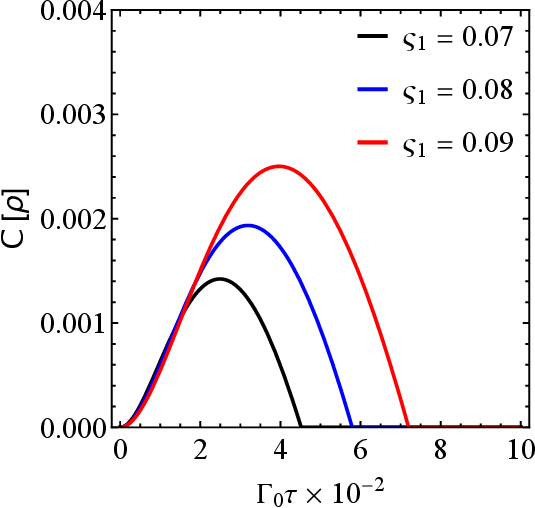}\label{harvg10}}\hspace*{2pt} 
	\subfloat[]{\includegraphics[height=2in]{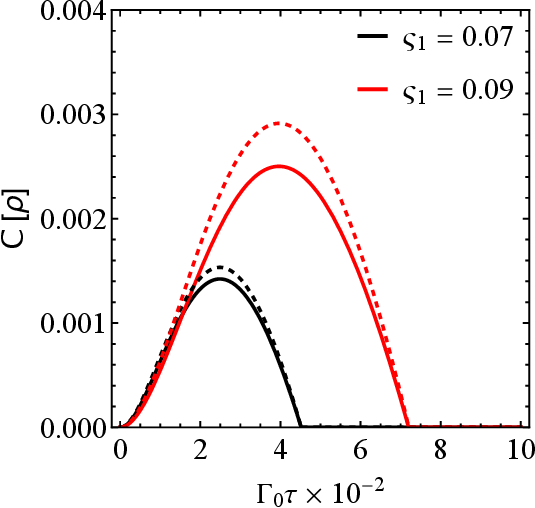}\label{harvg01}}	
	\caption{ Comparison of entanglement harvesting for different values of $ \varsigma_{1}$. The initial state were chosen to be $\ket{G}$ with $\varsigma_{2} = 0.58$. In (a) the atoms are polarized along the $x$ axis. In (b) we compare the concurrence for the system with the same initial configuration with the atoms polarized along different axis. The solid lines correspond to the atoms polarized along the $x$ axis and the dashed lines correspond to the atoms polarized along the $z$ axis.}
	\end{figure}

We now choose the initial state of the system as a separated state $\ket{G}$ or $\ket{E}$. In this scenario we wish to study the formation of entangled states. For the system prepared in the $\ket{G}$, both atoms in the ground state, entanglement can be generated from the excitation of the atoms due to the Unruh-Davies effect.  In Fig. \ref{harvg10} we observe this entanglement harvesting phenomenon when both atoms are polarized along the $x$ axis. We observe that, when entangled states are created, entanglement is increased by increasing $\varsigma_1$ (or decreasing the proper acceleration of the Atom $1$). We observe that for the values of $\varsigma_{\beta}$ used in the study of the entanglement sudden death, we do not obtain any entangled states.
In Fig. \ref{harvg01} we compare the concurrence for the atoms polarized along the $x$ axis (solid lines) with the atoms polarized along the $z$ axis (dashed lines). The systems are prepared in the same initial separated state $\ket{G}$ and same values of $\varsigma_1$ ans $\varsigma_2$. Now, we do not observe any significant difference in the entanglement lifetime, but we observe an increase in concurrence for atoms polarized along the $z$ axis.

\begin{figure}[htpb!]
    \centering
    \includegraphics[height=2in]{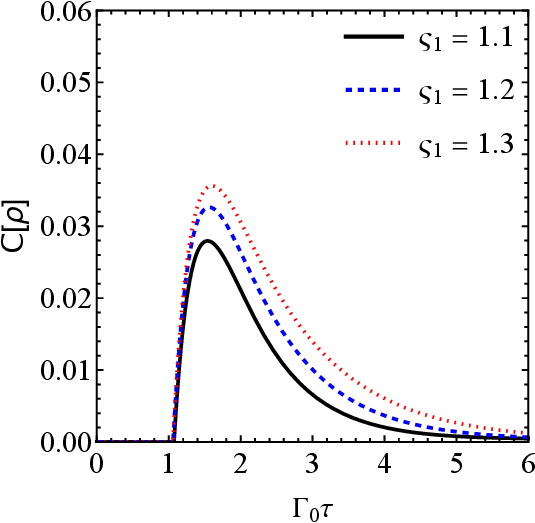}
    \caption{ Comparison of entanglement harvesting for different values of $\varsigma_{1}$. The initial state were chosen to be $\ket{E}$ with $\varsigma_{2} = 2.4$ and the polarization of the atoms along the $x-$direction.}
    \label{harve10}
    \end{figure}

For the initial state $\ket{E}$, entanglement can be created via spontaneous emission \cite{ficek2008delayed}. In this case, we only observe creation of entangled state for the polarization along the $x$ direction. For other polarizations the factor $2\sqrt{\rho_{GG}\rho_{EE}}$ is always bigger than $\sqrt{(\rho_{SS} - \rho_{AA})^2 -(\rho_{SA} - \rho_{AS})^2 }$. For  this initial separated state, the entanglement harvesting happens for a slightly large time interval $(\Gamma_{0} \tau)$ compared with our previous results and for some small choice of the acceleration parameters. This dynamics is shown in Fig. \ref{harve10}. It is important to point out that by increasing $\varsigma_1$ we are not able to use Eqs. (\ref{pgg}-\ref{peg}) anymore. Instead, we use the full form of the dissipative contribution given by Eq. (\ref{eqq:gme_em_field}).
\newpage

\begin{figure}[h!]
\centering
\includegraphics[height=2in]{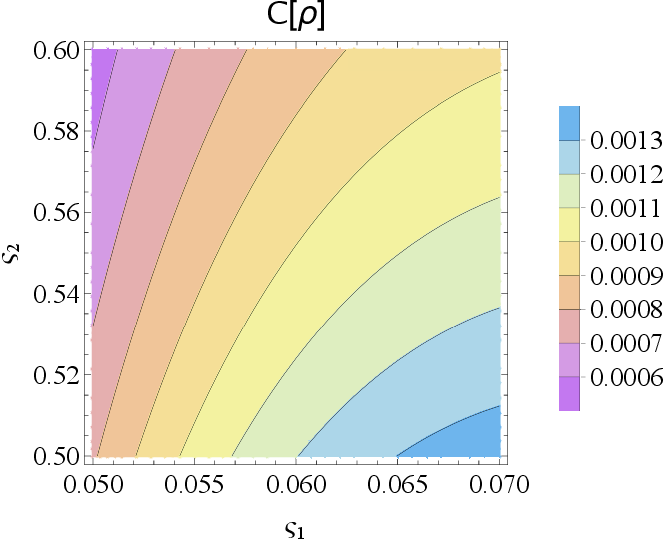}
\caption{Comparison of entanglement harvesting for different values of $\varsigma_{1}$ and $\varsigma_{2}$. The initial state were chosen to be $\ket{G}$ with $\Gamma_{0}\tau = 0.015$ and the polarization of the atoms along the $x$ direction.}
\label{harvboth}
\end{figure}

Finally, in Fig. \ref{harvboth} we observe how the change of both atoms' acceleration, $i.e.$, the change of $\varsigma_{1}$ and $\varsigma_{2}$, impacts the creation of entangled states. We have fixed $\Gamma_{0}\tau$ and set the atoms polarized along the $x$ direction. As argued before, we observe that the concurrence has a diversified dependence  of $\varsigma_1$ and $\varsigma_2$. If we fix the proper acceleration of the Atom $1$ and decrease the  proper acceleration of the Atom $2$, by increasing $\varsigma_{2}$, we observe that the concurrence decreases. Now, if we fix the proper acceleration of the Atom $2$, and decrease the acceleration of the Atom $1$, the concurrence increases. This pattern is also similar to the scalar field case \cite{soares23}. 
%
%
%
        \begin{table*}[htbp!]
    \caption{Comparison of the entanglement creation for uniformly accelerated atoms interacting with different type of Fields.}
    \label{comptable}
    \begin{tabular}{  l |  p{5cm} | p{5cm} }
        \toprule
     
& \textbf{Same proper acceleration's atoms}   
& \textbf{Different proper accelerations' atoms} \\\midrule
\textbf{Scalar Field} 
&   concurrence increases as the proper acceleration decreases \cite{benatti04_pra, hu2015}. Similar results can also be found in \cite{lin_2010, Salton_2015, liu2021does, wu2023birth, mann_2023correlation}. 
&  fixing the acceleration of one atom, the concurrence increases as the proper acceleration of the other atom decreases \cite{soares23}. \\\hline
\textbf{Electromagnetic Field}     
& \textbf entangled states are created and the concurrence increases as the proper acceleration decreases. For some choices of the orthogonal distance of the atom's, none entangled state were created for the atoms polarized along the $z$ direction \cite{yu2016pra}.                         
& concurrence increases as the proper acceleration of one atom decreases keeping the other atom's acceleration fixed. Entanglement creation does not happen for both atoms polarized along the $z$ direction with initial state $\ket{E}$.\\\hline
\textbf{Dirac Field }       
& entanglement states are created and the negativity decreases as the acceleration increases \cite{wu2023accelerating}.
& to be investigated. \\
        \bottomrule
    \end{tabular}
    \end{table*}

In Table \ref{comptable}, we summarize the differences/similarities between our results for entanglement creation with the results for entanglement dynamics in non-inertial frames for different quantum fields found in the literature \cite{benatti04_pra,lin_2010,hu2015, Salton_2015, liu2021does, wu2023birth, mann_2023correlation, soares23, yu2016pra, wu2023accelerating}.
\newpage
\section{Conclusions}\label{sec:conclusions}
Using the open quantum system framework, we investigate how the acceleration affects the entanglement dynamics of atoms in the scenario of Unruh-Davies effect. We have used the generalized master equation that allowed us to set the atoms with different values of proper accelerations. We would like to emphasize that we extended our previous result, where it was examined two-level systems coupled to a scalar field. Here we discuss the case of atoms interacting with an electromagnetic field. Although in Ref. \cite{yu2016pra} the authors discussed atoms interacting with an electromagnetic field, only the case of same proper acceleration was studied. Here we are assuming that the atoms follow different hyperbolic trajectories with different proper times, which introduces some technical difficulties. As we are considering a pair of electric dipoles, polarization effects of atoms were also discussed.

By considering an initial non-separable state for the atoms and changing the acceleration and polarization conditions, we shed light on the behavior and degradation of initial entangled states. We observe that once the system is prepared in an entangled state different polarization choices lead us to enlarge the lifetime of the entangled state. This happens for both state $\ket{S}$ and state $\ket{A}$. In this phenomenon, for a fixed value of the proper acceleration of one atom, we observe that increasing the proper acceleration of the other decreases the lifetime of the entangled state. The concurrence when both accelerations vary was also studied.

We also explore the phenomena of entanglement harvesting by considering an initial separable state for the atoms. In this case, we observe that for some polarization of the atoms we do not observe any creation of entangled states in some initial states of the system. Entangled states were only created when we significantly reduced the atoms' proper accelerations. Preparing the system in the state $\ket{G}$, fixing the proper acceleration of the Atom $2$ and varying the proper acceleration of the Atom $1$, we observed that entangled states were created and the concurrence increases when the proper acceleration, or $\varsigma_{1}$, decreases. For the system prepared in the state $\ket{E}$ and both atoms polarized along the $x$ we observe a similar entanglement dynamics. The polarization of the atoms plays a crucial role in this case since we do not observe entanglement harvesting for the polarization along the $y$ or $z$ direction.

One natural continuation is to discuss if the direction of the proper acceleration alters our results using the formalism derived in Ref. \cite{soares23}. See also, 
for example, the Ref. \cite{liu2021does}. This work also can be used to discuss the dynamics of entanglement with atoms in a curved spacetime. For instance, near the event horizons of a non-rotating black hole, one can use the Rindler coordinates to discuss entanglement degradation and creation phenomena in the Schwarzschild metric \cite{caribe23}. This topic is under investigation by the authors.
\section*{Acknowledgments}

This work was partially supported by Conselho Nacional de Desenvolvimento Cient\'{\i}fico e Tecnol\'{o}gico -- CNPq, under grants 303436/2015-8 (N.F.S.) and 317548/2021-2 (G.M.),~and Funda\c{c}\~ao Carlos Chagas Filho de Amparo \`a Pesquisa do Estado do Rio de Janeiro -- FAPERJ under grants E-26/202.725/2018 and E-26/201.142/2022 (G.M.). 

\appendix
\section{Computation of the $\Gamma^{(\alpha \beta)}$ function}\label{app:gammas}

In this appendix we present the explicit computation of the function $\Gamma_{ij}^{(\alpha \beta)}(a_{\beta} \omega)$. To do so, we start using Eq. (\ref{eqq:corre_final_form}) into (\ref{eqq:gammaij}). Therefore, our function is written as the following
    \begin{eqnarray}
        \Gamma_{ij}^{(\alpha \beta)}(a_{\beta} \omega) =  \frac{a^{4} \delta_{ij}}{4 \pi^{2} a_{\alpha} a_{\beta} } \int_{-\infty}^{\infty} ds~e^{i a_{\beta} \omega s} \frac{\nu_{i}^{(\alpha \beta)}(s)}{\left[\cosh \left(as - i  \epsilon\right) - \cosh\phi\right]^{3}},\label{app:wij_general}
    \end{eqnarray}
with
\begin{eqnarray}
\nu_{1}^{(\alpha \beta)}(s) &=& \cosh(as - i \epsilon) - \cosh\phi,\\
\nu_{2}^{(\alpha \beta)}(s)  &=& \nu_3^{(\alpha \beta)}(s)  = \cosh(as - i \epsilon)\cosh\phi - 1. 
\end{eqnarray}
For $\alpha=\beta$, we have $\cosh \phi = 1$ and $\nu_{1}^{(\beta \beta)}(s) = \nu_{2}^{(\beta \beta)}(s) = \nu_{3}^{(\beta \beta)}(s) = 2\sinh^{2}(as/2)$. The above equation can be simplified as:
\begin{equation}
\Gamma_{ij}^{\beta \beta}(a_{\beta}\omega)= \frac{a^{4} \delta_{ij} }{16 \pi^{2}  a_{\beta}^{2} } \int_{-\infty}^{\infty} ds~\frac{e^{i a_{\beta} \omega s}}{\sinh^{4}\left(\frac{as}{2} - i \epsilon\right)}.\label{app:gamaii}
\end{equation}
The integral that appeared in Eq. (\ref{app:gamaii}) is similar to the one that appears in the scalar case and has the solution \cite{soares23, yu2016pra}
	\begin{eqnarray}
		\Gamma^{(\beta \beta)}_{ij}(a_{\beta}\omega) = \frac{\delta_{ij} a_{\beta}  \omega^3}{3\pi} \left(1 + \frac{a^2}{a_{\beta}^{2}\omega^{2}}\right)\coth\left(\frac{\pi a_{\beta}\omega}{a}\right). \label{app:gamma_aa} 
	\end{eqnarray}
For $\alpha \neq \beta$ and $i = j = 1$ we have to solve the following integral
    \begin{eqnarray}
        \Gamma_{11}^{(\alpha \beta)}(a_{\beta} \omega) =  \frac{a^{3} }{4 \pi^{2} a_{\alpha} a_{\beta} } \int_{-\infty}^{\infty} du~ \frac{e^{ \frac{i a_{\beta} \omega u}{a} }}{\left[\cosh \left(u - i \epsilon\right) - \cosh\phi\right]^{2}},\label{app:gamma_11_general}
    \end{eqnarray}
where we again have defined $u = as$. We then perform the contour deformation $u = r - i \pi$ and Eq. (\ref{app:gamma_11_general}) becomes
    \begin{eqnarray}
        \Gamma_{11}^{(\alpha \beta)}(a_{\beta} \omega) =  \frac{a^{3}e^{\frac{a_{\beta} \omega \pi}{a}} }{4 \pi^{2} a_{\alpha} a_{\beta} } \int_{-\infty}^{\infty} dr~ \frac{\cos\left(\frac{a_{\beta} \omega r}{a}\right) + i \sin\left(\frac{a_{\beta} \omega r}{a}\right)}{\left[\cosh \left(r\right) + \cosh\phi\right]^{2}}.\label{app:gamma_11_rchange}
    \end{eqnarray}	

\noindent We observe that the imaginary part of the integral of Eq. (\ref{app:gamma_11_rchange}) vanishes and we can write
    \begin{eqnarray}
        \Gamma_{11}^{(\alpha \beta)}(a_{\beta} \omega) =  \frac{a^{3}e^{\frac{a_{\beta} \omega \pi}{a}} }{2 \pi^{2} a_{\alpha} a_{\beta} } \int_{0}^{\infty} dr~ \frac{\cos\left(\frac{a_{\beta} \omega r}{a}\right)}{\left[\cosh \left(r\right) + \cosh\phi\right]^{2}}.\label{app:gamma_11_rchange_f}
    \end{eqnarray}	
The real part can be solved by using \cite{gradshteyn2014table}
	\begin{equation}
	\int_{0}^{\infty}\frac{\cos ax}{b \cosh  x + c}dx = \frac{\pi \sin\left(a~{arccosh}\frac{c}{b}\right)}{\sqrt{c^2 - b^2} \sinh a\pi},\label{app:gradstein_result}
	\end{equation}
where we observe that Eq. (\ref{app:gamma_11_rchange_f}) can be written using that
	\begin{equation}
	-\frac{\partial}{\partial c}\int_{0}^{\infty}\frac{\cos ax}{b \cosh  x + c}dx = \int_{0}^{\infty}\frac{\cos ax}{\left(b \cosh  x + c\right)^{2}}dx.
	\end{equation}
Therefore, the integral of Eq. (\ref{app:gamma_11_rchange_f}) can be solved using the derivative of (\ref{app:gradstein_result}) with respect to $c$ and then using $a = \frac{a_{\beta} \omega}{a}$, $b = 1$ and $c = \cosh \phi$. After these manipulations one obtains
	\begin{eqnarray}
	  \Gamma_{11}^{(\alpha \beta)}(a_{\beta} \omega) = \frac{a^{3}}{\pi a_{\alpha} a_{\beta}} \frac{{csch}^{2} \phi}{1 - e^{-2 \pi \frac{ a_{\beta} \omega }{a}}}\left[\coth \phi \sin\left(\frac{a_{\beta} \omega \phi}{a}\right) - \frac{a_{\beta} \omega }{a}\cos\left(\frac{a_{\beta} \omega \phi}{a}\right)\right].\label{app:gamma1_final}
	\end{eqnarray}
For $\alpha \neq \beta$ and $i = j = 2$ or $3$ we have
	\begin{eqnarray}
	  \Gamma_{22}^{(\alpha \beta)}(a_{\beta} \omega) = \Gamma_{33}^{(\alpha \beta)}(a_{\beta} \omega) = \frac{a^{4}}{4 \pi^{2} a_{\alpha} a_{\beta}}\int_{-\infty}^{\infty} ds~ e^{i a_{\beta} \omega s}\frac{ \left(\cosh (as - i \epsilon) \cosh \phi - 1\right)}{(\cosh (as - i \epsilon) - \cosh \phi)^{3}}.\label{app:gamma22_gen}
	\end{eqnarray}
Following the same manipulations we have used for $\Gamma_{11}^{(\alpha \beta)}$ we obtain, after the contour deformation $u = r - i \pi$, the following integrals:
	\begin{eqnarray}
	\Gamma_{22}^{(\alpha \beta)}(a_{\beta} \omega) = \frac{a^{4}}{2 \pi^{2} a_{\alpha} a_{\beta}}\left[\cosh \phi\int_{0}^{\infty} dr~ \frac{ \cos(\frac{a_{\beta} \omega}{a} r)\cosh r }{(\cosh r + \cosh \phi)^{3}}\right. \nonumber\\ 
	\left. + \int_{0}^{\infty} dr~ \frac{ \cos(\frac{a_{\beta} \omega}{a} r)}{(\cosh r + \cosh \phi)^{3}}\right].\label{app:gamma22_genf}
	\end{eqnarray}
To solve the above integrals we can also use Eq. (\ref{app:gradstein_result}). Using $a = \frac{a_{\beta} \omega}{a}$, $b = 1$ and $c = \cosh \phi$ we can write
	\begin{eqnarray}
	\frac{1}{2}\frac{\partial^{2}}{\partial c\partial b}\int_{0}^{\infty}\frac{\cos ax}{b \cosh  x + c}dx &=& \int_{0}^{\infty}\frac{\cos ax \cosh x}{(b \cosh  x + c)^3}dx,\label{eqq:grad_1}\\
	\frac{1}{2}\frac{\partial^{2}}{\partial c^{2}}\int_{0}^{\infty}\frac{\cos ax}{b \cosh  x + c}dx &=& \int_{0}^{\infty}\frac{\cos ax }{(b \cosh  x + c)^3}dx.\label{eqq:grad_2}
	\end{eqnarray}
Performing the aforementioned derivatives of Eq. (\ref{app:gradstein_result}), we obtain the following result of Eq. (\ref{app:gamma22_genf}) as
	\begin{eqnarray}
	 \Gamma_{22}^{(\alpha \beta)}(a_{\beta} \omega) = \Gamma_{33}^{(\alpha \beta)}(a_{\beta} \omega) = \frac{a}{2 \pi a_{\alpha} a_{\beta}} \frac{{csch}^{2} \phi}{1 - e^{-2 \pi \frac{ a_{\beta} \omega }{a}}} \left[\frac{a_{\beta} \omega}{a} \cos \right(\frac{a_{\beta} \omega}{a} \phi\left) \cosh \phi\right.\nonumber\\ \left. + \sin\left(\frac{a_{\beta} \omega}{a}\phi\right)\sinh \phi\left(\left(\frac{a_{\beta} \omega}{a}\right)^{2} - {csch}^{2} \phi\right)\right].\label{app:gamma2_final}
	\end{eqnarray}
For simplicity we can group Eqs. (\ref{app:gamma1_final}) and (\ref{app:gamma2_final}) as
\begin{equation}
\Gamma^{(\alpha \beta)}_{ij}(a_{\beta} \omega) = \frac{a^{3}}{4 \pi a_{\alpha} a_{\beta}}\frac{{csch}^{2}\phi}{1 - e^{-2 \pi \frac{ a_{\beta} \omega }{a}}}~\delta_{ij}\mathcal{U}_{i}\left(a_{\beta} \omega\right),
\end{equation}
where we have defined the functions
	\begin{eqnarray}
\mathcal{U}_{1}\left(\frac{a_{\beta} \omega}{a}\right) &=& 4\left[\coth \phi \sin\left(\frac{a_{\beta} \omega \phi}{a}\right) - \frac{a_{\beta} \omega }{a}\cos\left(\frac{a_{\beta} \omega \phi}{a}\right)\right],\label{u1}\\
\mathcal{U}_{2}\left(\frac{a_{\beta} \omega}{a}\right) &=& \mathcal{U}_{3}\left(\frac{a_{\beta} \omega}{a}\right) \nonumber\\  &=& 2\left[\frac{a_{\beta} \omega}{a} \cos \right(\frac{a_{\beta} \omega}{a} \phi\left) \cosh \phi + \sin\left(\frac{a_{\beta} \omega}{a}\phi\right)\sinh \phi\left(\left(\frac{a_{\beta} \omega}{a}\right)^{2} - {csch}^{2} \phi\right)\right].\label{u2}
	\end{eqnarray}

It is important to remember that we are not considering the summation convention over repeated indices in this work.

\section{Explicit form of the generalized master equation}\label{app:gmeterms}

In this appendix we present the explicit form of the master equation where we consider the contribution only of the dissipative part of it as discussed in Sec. \ref{sec:ent_dyn}. From Eq. (\ref{eqq:rho_initial_basis}), considering $\varsigma_{1} << \varsigma_{2}$,  we only have to compute eight components that are shown in the following

	\begin{eqnarray}
	\frac{d}{d\tau}\rho_{GG}(\tau) &=& (-2 A^{(11)} + 2 B^{(11)}) \rho_{GG}(\tau)\nonumber\\ &+& (A^{(11)} + B^{(11)} - 2 (B^{(12)} - B^{(22)} + B^{(21)})) \rho_{AA}(\tau)\nonumber\\ &+& (A^{(11)} + B^{(11)} - 2 (B^{(12)} + B^{(22)} -  B^{(21)})) \rho_{AS}(\tau)\nonumber\\  &+&(A^{(11)} + B^{(11)} + 2 B^{(12)} - 2 B^{(22)} - 2 B^{(21)}) \rho_{SA}(\tau)\nonumber\\ &+& (A^{(11)} + B^{(11)} + 2 B^{(12)} + 2 B^{(22)} + 2 B^{(21)}) \rho_{SS}(\tau),\label{pgg}
	\end{eqnarray}
	\begin{eqnarray}
\frac{d}{d\tau}\rho_{SS}(\tau) &=& (A^{(11)} - B^{(11)}) \rho_{GG}(\tau)\nonumber\\ 
&+& (-B^{(11)} + B^{(12)} + B^{(22)} - B^{(21)}) \rho_{AS}(\tau)\nonumber\\ &+& (A^{(11)} + B^{(11)} + 2 (B^{(12)} + B^{(22)} + B^{(21)})) \rho_{EE}(\tau)\nonumber\\ &+& (-B^{(11)} - B^{(12)} + B^{(22)} + B^{(21)}) \rho_{SA}(\tau)\nonumber\\ &+& (-2 A^{(11)} - 2 B^{(12)} - 2 B^{(22)} - 2 B^{(21)}) \rho_{SS}(\tau),\label{pss}
	\end{eqnarray}
	\begin{eqnarray}
\frac{d}{d\tau}\rho_{AA}(\tau) &=& (A^{(11)} - B^{(11)}) \rho_{GG}(\tau)\nonumber\\ &-&2 (A^{(11)} - B^{(12)} + B^{(22)} - B^{(21)}) \rho_{AA}(\tau)\nonumber\\  &+& (-B^{(11)} + B^{(12)} + B^{(22)} - B^{(21)}) \rho_{AS}(\tau)\nonumber\\  &+& (A^{(11)} + B^{(11)} - 2 B^{(12)} + 2 B^{(22)} - 2 B^{(21)}) \rho_{EE}(\tau)\nonumber\\   &+& (-B^{(11)} - B^{(12)} + B^{(22)} + B^{(21)}) \rho_{SA}(\tau),\label{paa}
	\end{eqnarray}
	\begin{eqnarray}
\frac{d}{d\tau}\rho_{EE}(\tau) &=& -2 ( A^{(11)} + B^{(11)} +2 B^{(22)}) \rho_{EE}(\tau)\nonumber\\ &+&(A^{(11)} - B^{(11)}) \rho_{AA}(\tau)\nonumber\\  &+& (-A^{(11)} + B^{(11)}) \rho_{AS}(\tau)\nonumber\\   &+& (-A^{(11)} + B^{(11)}) \rho_{SA}(\tau)\nonumber\\  &+& (A^{(11)} - B^{(11)}) \rho_{SS}(\tau),\label{pee}
	\end{eqnarray}
	\begin{eqnarray}	
\frac{d}{d\tau}\rho_{AS}(\tau) &=&  (A^{(11)} - B^{(11)}) \rho_{GG}(\tau)\nonumber\\ &-& 2 (A^{(11)} + B^{(22)}) \rho_{AS}(\tau)\nonumber\\  &+&(-B^{(11)} - B^{(12)} + B^{(22)} + B^{(21)}) \rho_{AA}(\tau)\nonumber\\   &+& (-A^{(11)} - B^{(11)} + 2 B^{(12)} + 2 B^{(22)} - 2 B^{(21)}) \rho_{EE}(\tau)\nonumber\\   &+& (-B^{(11)} - B^{(12)} + B^{(22)} + B^{(21)}) \rho_{SS}(\tau),\label{pas}
	\end{eqnarray}
	\begin{eqnarray}
\frac{d}{d\tau}\rho_{SA}(\tau) &=&  (A^{(11)} - B^{(11)}) \rho_{GG}(\tau)\nonumber\\  &+&(-B^{(11)} + B^{(12)} + B^{(22)} - B^{(21)}) \rho_{AA}(\tau)\nonumber\\  &+& (-A^{(11)} - B^{(11)} - 2 B^{(12)} + 2 B^{(22)} + 2 B^{(21)}) \rho_{EE}(\tau)\nonumber\\  &-&2 ( A^{(11)} + B^{(22)}) \rho_{SA}(\tau)\nonumber\\  &+& (-B^{(11)} + B^{(12)} + B^{(22)} - B^{(21)})\rho_{SS}(\tau),\label{psa}
	\end{eqnarray}
	\begin{eqnarray}
\frac{d}{d\tau}\rho_{GE}(\tau) = -2 (A^{(11)} + B^{(22)}) \rho_{GE}(\tau),\label{pge}\\
\frac{d}{d\tau}\rho_{EG}(\tau) = -2 (A^{(11)} + B^{(22)}) \rho_{EG}(\tau).\label{peg}
	\end{eqnarray}
\bibliographystyle{iopart-num}
\bibliography{references}
\end{document}